\documentclass[aps,prx,twocolumn,floats,showpacs,superscriptaddress,nofootinbib]{revtex4-1}
\usepackage{graphicx,epsfig}
\usepackage{times,bbm}
\usepackage{graphics,dcolumn,bm,float}
\usepackage{amssymb,amsmath,rotate,color,amsfonts}
\usepackage[title,titletoc,toc]{appendix}
\usepackage{mathtools}
\usepackage{booktabs}
\usepackage{tcolorbox}
\usepackage[pagebackref=false,colorlinks,linkcolor=magenta,citecolor=blue,urlcolor=magenta]{hyperref}

\usepackage{wrapfig}
\usepackage{lipsum}
\usepackage{mwe}
\usepackage[mathlines]{lineno}
\usepackage[mathscr]{euscript}
\usepackage{hyperref}
\usepackage{breqn}
\usepackage{bbold}
\usepackage{pgfplots}
\usepackage{float}
\usepackage{tkz-euclide}
\usepackage{braket}
\usepackage{physics}
\usepackage[caption=false]{subfig}
\usepackage[export]{adjustbox}
\usepackage{tikz}
\usepackage{slashed}
\usepackage{bm}
\usetikzlibrary{through,calc}
\usetikzlibrary{positioning}

\newcommand{\hlt}[1]{\textcolor{black}{#1}}

\newcommand{\be}{\begin{equation}}
\newcommand{\ee}{\end{equation}}

\newcommand{\bea}{\begin{eqnarray}}
\newcommand{\eea}{\end{eqnarray}}
\newcommand{\nn}{\nonumber}

\newcommand{\bs}{\vec}
\newcommand{\bmt}{\left[\begin{matrix}}
\newcommand{\emt}{\end{matrix}\right]}

\begin{document}
\preprint{}
\title{\hlt{Moving frame} theory of zero-bias photocurrent \hlt{on the surface of} topological insulators}

\author{S. A. Jafari}
\email{akbar.jafari@rwth-aachen.de}
\affiliation{Peter Gr\"unberg Institute 2$,$ Forschungszentrum J\"ulich$,$ 52425 J\"ulich$,$ Germany}
\date{\today}

\begin{abstract}
Motivated by observations of zero-biased photocurrent on the surface of topological insulators, we show that the in-plane effective magnetic field $\tilde B$ implements a moving frame transformation on the topological insulators' helical surface states. As a result, photo-excited electrons on the surface undergo a Galilean boost proportional to the effective in-plane magnetic field $\tilde B$. The boost velocity is transversely proportional to $\tilde B$. This explains why the experimentally observed photocurrent depends linearly on $\tilde B$. Our theory while consistent with the observation that at leading order the effect does not depend on the polarization of the incident radiation, at next leading order in $\tilde B$ predicts a polarization dependence in both parallel and transverse directions to the polarization. We also predict two induced Fermi surface effects that can serve as further confirmation of our moving frame theory. \hlt{Based on the estimated value $\zeta\approx 0.34$ of the tilt parameter for a magnetic fields of $\tilde B\sim 3$T, our} geometric picture qualifies the surface Dirac cone of magnetic topological insulators as an accessible platform for the synthesis and experimental investigation of \hlt{strong synthetic gravitational phenomena}. 
\end{abstract}



\maketitle
\narrowtext

\section{Introduction}
Surface states of topological insulators (TIs)~\cite{Hasan2010} feature a helical Dirac cone~\cite{Shen2017} where the two-dimensional Pauli matrices representing the associated Clifford algebra are realized by the actual spin of the electrons. This simple fact means that external magnetic influence can be employed to manipulate them in interesting ways. Ideally their Hamiltonian is given by $v(p_x\sigma_y-p_y\sigma_x)$. Breaking the time-reversal (TR) invariance by a magnetic field perpendicular to the surface gives rise to a $m\sigma_z$ term that opens up a gap, whereas in-plane magnetic field being a combination of $\sigma_x$ and $\sigma_y$ terms only shifts the location of the Dirac node, but does not open up a gap. 
The in-plane fields give rise to fascinating effects such as planar Hall effect~\cite{Taskin2017,Zheng2020}. 
It has been experimentally found that in presence of in-plane magnetic field, a zero-bias photo-current 
flows in the sample whose direction is transverse to the in-plane magnetic field, and its magnitude is 
proportional to the amount of the in-plane magnetic field~\cite{Ogawa2016}. A similar photo-current is observed at the zero external magnetic field~\cite{Kuroda2016}. This effect is a basis for the application of TIs in infrared optoelectronics~\cite{Kiemle2020}. The fact that $\sigma$ matrices in the surface Dirac cone of TIs are actual spins, \hlt{makes it plausible} to use magnetic fields or magnetization to influence the Dirac cones. Can the Dirac cones be manipulated in other interesting ways? In this paper by providing a theory for the observed zero-bias photocurrent \hlt{on the surface of} TIs, we show that surface Dirac cones in TIs are promising platform to manipulate Dirac cones in ways that resembles the the way gravitational sources influences the null geodesics (i.e. cones)~\cite{Davis2022}. 

The above simple theory of helical Dirac cones ignores an important fact: The bulk band edge states have two completely different characters with opposite parities~\cite{Shen2017}, such as $s$ and $p$ character. As such, the particle-hole (PH) asymmetry is a genuine property of the helical Dirac cones as is evident from photemission data both in bulk~\cite{Hasan2010,Hasan2011} thin films of TIs~\cite{Zhang2010,Vidal2013} and must be incorporated into theoretical considerations~\cite{Shan2010}. The PH symmetry is represented by the fact that Pauli matrices $\sigma_{x,y}$ that construct the helical Dirac states anti-commute with $\sigma_z$. The way to break PH symmetry is to add a term \hlt{that does not anti-commute with $\sigma_z$ and hence it must be} proportional to $\sigma_0$, unit $2\times 2$ matrix.  A constant term proportional to $\sigma_0$ is ruled out as it redefines the energy axis. A linear term in $\vec p$ can be of the form $\vec\zeta^{(0)}.\vec p~\sigma_0$ with the provision that since it is odd in $\vec p$, the other Dirac cone on the other surface must have the same form with opposite $\vec\zeta^{(0)}$ value. There can also be a term even in $\vec p$ that can be combined to give,
\begin{equation}
       H_0= v(p_x\sigma_y-p_y\sigma_x)+\frac{p^2}{2m_*}\sigma_0+\tau_z\vec\zeta^{(0)} \vec p.\sigma_0
   \label{PHA.eqn}
\end{equation}
where $\tau_z=\pm 1$ correspond to top and bottom surface and the PH asymmetry is represented by $m_{*}^{-1}$. The PH symmetric situation corresponds to $m_{*}^{-1}\to 0$. Quadratic PH breaking term also appears in a microscopic derivation of the TI surface states~\cite{Liu2010} as well as in TI thin films~\cite{Lu2010}. Note that we are ignoring the trigonal warping terms~\cite{Fu2009,Chen2009} that are \hlt{negligible in the experiment of Ref.~\cite{Ogawa2016}. }

The purpose of this letter is two-fold: (1) To construct a theory the \emph{zero bias} photo-current in magnetic~\cite{Ogawa2016} and non-magnetic TIs~\cite{Kuroda2016} that relies on PH asymmetry and time-reversal (TR) symmetry for the observed zero-bias photo-current response with additional higher order predictions. (2) To reveal a moving frame picture of Galilean boost inherent in the Hamiltonian~\eqref{PHA.eqn} \hlt{once it is subjected into an external in-plane magnetic influence}. The latter establishes an intriguing paradigm of synthesizing a large class of emergent spacetime structures via engineering of the spin-orbit coupled Dirac cones by external magnetic influence. 

\section {In-plane magnetic field as knob to tune tilting of Dirac cone}
Let us couple external magnetic field $\vec B$ to Eq.~\eqref{PHA.eqn}. This magnetic field in Ref.~\cite{Ogawa2016} is denoted by $B_y$. As can be seen in Fig.~3a,b of Ref.~\cite{Ogawa2016}, even at temperatures above the Curie temperature $T_c=85$K of the Cr atoms, the photo-current continues to be linear in $\vec B$. For $T>T_c$, the magnetization $\vec m$ of the Cr atoms is zero and the entire photo-current is caused by the external $\vec B$ field. This indicates that both the Direct Zeeman coupling to external $\vec B$ and exchange coupling to magnetization $\vec m$ are important in the response of the Dirac cone. Therefore we introduce an effective magnetic field $\tilde B=\vec B+\vec m$ where the magnetization $\vec m=\vec m(T,\vec B)$ of the Cr atoms is function of temperature and external magnetic field. 

The Hamiltonian in the presence of $\tilde B$ and  PH asymmetry will become
\begin{equation}
   H_0= v(p_x\sigma_y-p_y\sigma_x)+\frac{p^2}{2m_{*}}\sigma_0+\tau_z\vec\zeta^{(0)}.\vec p~\sigma_0-\hlt{g}\mu_B\tilde B.\vec\sigma
\end{equation}
where $\mu_B$ is the Bohr magneton \hlt{and $g$ is the g-factor that can be large due to spin-orbit coupled nature of the bands~\cite{Wolos2016}}.
A magnetic field along the $z$ direction will give a $\sim \tilde B_z\sigma_z$ term opening a gap. But we would like  to understand the role of in-plane magnetic field components $\tilde B_a$ with $a=1,2$ for $x,y$ directions, respectively. Upon coupling a planar B-field, not only the location of the Dirac node will be shifted. 
Such a shifting of Dirac cones is supported by ab-initio calculations for Cr-doped Bi$_2$Se$_3$ even when the external field $B$ is zero~\cite{Kim2016}. 
But driven by PH asymmetry, additional terms will also be generated as follows: By shifting the momenta according to
\begin{equation}
   p_a\to p_a+\hlt{g}\frac{\mu_B}{v}\varepsilon_{ab}\tilde B_b,~~~a,b=1,2
\end{equation}
where $\varepsilon_{ab}$ is totally ant-isymmetric tensor in planar indices $1,2$ with $\varepsilon_{12}=1$, we obtain
\begin{eqnarray}
   &&H_0 = v(p_x\sigma_y-p_y\sigma_x)+\frac{p^2}{2m_{*}}\sigma_0
   +\tau_z\vec\zeta^{(0)}.\vec p ~\sigma_0\nn\\
   &&+\frac{\hlt{g}\mu_B}{m_*v}(p_x\tilde B_y-p_y\tilde B_x)\sigma_0+\frac{(\hlt{g}\mu_B)^2}{2m_*v^2}(\tilde B_x^2+\tilde B_y^2)\sigma_0,
   \label{tiltEM.eqn}
\end{eqnarray}
where the terms in the second line are  generated by shifting the Dirac node. \hlt{The first, second and fourth terms in the above equation are used in Ref.~\cite{Zheng2020} 
to explain the planar Hall effect, whearas} the last term in the second line is a shift in the chemical potential, but caused by in-plane magnetic field~\cite{Hama2018}. 
\hlt{In fact our moving frame interpretation can be identified as underlying mechanism for planar Hall effect~\cite{Moradpouri2023} as well as the electric-field effects
induced by magnetic influences simply becasuse in a moving frame}  magnetic field appears to have an effect like an electric field. \hlt{In addition this interpretation
explains the zero-bias photocurrent induced by magnetic field.} The \hlt{latter} becomes clear soon. Therefore from Eq.~\eqref{tiltEM.eqn} we obtain a magnetically driven 
tilting and chemical potential shift:
\begin{equation}
   \zeta^{(1)}_a=\varepsilon_{ab}\frac{\hlt{g}\mu_B \tilde B_b}{m_*v^2},~~~a,b=1,2,~~~~\tilde\mu=\frac{|\zeta^{(1)}|\hlt{g} \mu_B|\tilde B|}{2}.
   \label{zetaB.eqn}
\end{equation}
The first order term in $\tilde B$ is $v\vec\zeta^{(1)}.\vec p~\sigma_0$ is responsible for tunability of the tilting and magnetic-field induced gating of the Dirac cone
at the surface of topological insulators. The intrinsic (zeroth order) tilt $\tau_z\vec\zeta^{(0)}$ that may exist even in the absence of $\tilde B$ can be \hlt{either} zero or non-zero. As we will see shortly, the data in both magnetic and non-magnetic TIs support a non-zero $\bs\zeta^{(0)}$.  Let us start by formulation of the linear term. The second order (in $\tilde B$) term is a magnetic induced chemical potential. The above analysis shows that an in-plane magnetic field can be used to tune the tilting in the surface Dirac cone of topological insulators.  In this way the total tilt will have a constant \hlt{(background)} term $\tau_z\bs\zeta^{(0)}$ and a \hlt{tilt} term \hlt{$\zeta^{1}$} linear in the effective in-plane magnetic field as in~\eqref{zetaB.eqn}. 

\section{Tilt term as moving frame} 
To see the physics arising from the tilting of the Dirac cone, let us for the moment forget the $\tilde B$-induced chemical potential in Eq.~\eqref{tiltEM.eqn}. Representing the total tilt as $\vec\zeta=\vec\zeta^{(0)}+\vec\zeta^{(1)}$,  the Hamiltonian will be $v(p_x\sigma_y-p_y\sigma_x)+v\vec\zeta.\vec p~\sigma_0$ whose eigenvalues are $\varepsilon_\pm=v\vec\zeta.\vec p\pm v p$ where $p=|\vec p|$ is the magnitude of momentum.  This is the dispersion relation of a Dirac cone which is tilted along vector $\vec \zeta$ by an amount proportional to dimensionless number $\zeta$.  Already at this level one can immediately see that the parameter $\zeta$ changes the \hlt{speed} of right/left movers \hlt{by opposite amounts}. In order to see this clearly, consider one spatial dimension $x$ where $\varepsilon_\pm=v(\zeta p_x\pm p_x)$. This means that the right moving electrons' velocity is boosted as $(\zeta+1)v$, while the left moving electrons' velocity is $(1-\zeta)(-v)$.  Such a {\em velocity asymmetry} between right and left movers is \hlt{naturally} equivalent to viewing the movement of Dirac electrons in a frame moving with velocity $-v\vec\zeta$. \hlt{The non-trivial fact is that such a moving frame effect can be generated by an in-plane magnetic field as in Eq.~\eqref{zetaB.eqn}. Even more non-trivial fact is that as we will see in this paper, the fraction $\zeta$ can be as large as $1/3v$ (where the Fermi velocity $v$ is the upper limit of speeds in the material) the mechanical analogue of which would  be to imagine a frame moving with a third of speed of light.}

To see how such a boost on Dirac electrons gives rise to a non-trivial spacetime geometry, let us combine the energy and momentum to define 1+2 dimensional covariant momentum $p_\mu=(-\varepsilon/v,p_x,p_y)$. Then taking the $\vec\zeta.\vec p$ term to the left side of the dispersion relation and squaring both sides (upon which $\pm$ disappears), the resulting equation being quadratic form in $p_\mu$ can be written as~\cite{Volovik2021}
\begin{equation}
   g^{\mu\nu}p_\mu p_\nu=0,
\end{equation}
where the contra-variant components of the {\em emergent spacetime metric} can be immediately read off:
\begin{equation}
   g^{\mu\nu}=\begin{pmatrix}
   -1		&-\zeta_x	&-\zeta_y\\
   -\zeta_x	&1-\zeta_x^2	&-\zeta_x\zeta_y\\
   -\zeta_y	&-\zeta_x\zeta_y&1-\zeta_y^2
   \end{pmatrix}.
   \label{metriccontra.eqn}
\end{equation}
The inverse of the above metric gives the covariant components
\begin{equation}
   g_{\mu\nu}=\begin{pmatrix}
   -1+\zeta^2	&-\zeta_x	&-\zeta_y\\
   -\zeta_x	&1		&0\\
   -\zeta_y	&0		&1
   \end{pmatrix},
   \label{metriccovar.eqn}
\end{equation}
where $\zeta^2=\zeta_x^2+\zeta_y^2$. The compact representation of Eq.~\eqref{metriccovar.eqn} is
\begin{equation}
   ds^2=-(vdt)^2+(d\vec x-\vec\zeta v dt)^2. 
   \label{invariantdistance.eqn}
\end{equation}

For the upright Dirac cone where there is no left/right velocity asymmetry, the tilt parameter is $\zeta=0$ and the invariant spacetime distance is $ds^2=-(vdt)^2+(d\vec x)^2$.  The relation between the latter and the invariant distance~\eqref{invariantdistance.eqn} in presence of a tilt is given by the Galilean transformation $(vt,\vec x)\to (v t,\vec x-\vec\zeta vt)$, where the boost parameter $\vec\zeta$ according to Eq.~\eqref{zetaB.eqn} is controlled by the effective in-plane magnetic field.  \emph{Therefore the surface states of topological insulators subjected to in-plane magnetic  field are solid-state realizations of moving frame where the frame velocity $-v\bs\zeta$ is tunable by in-plane magnetic field.}  Larger effective in-plane magnetic fields $\tilde B$ corresponds to faster moving frames. The direction of movement of the frame is always transverse to $\tilde B$. 

Our moving frame interpretation is the most natural way in which optical absorption becomes a source of photo-current. In the absence of tilt, the states in the upper branch of the Dirac cone are populated by the absorption of light. But since left/right mover's velocities are symmetric, there will be no net current. In this way the photo-current is cancelled out and remains inert. However, in presence of tilting, right and left mover's velocity will not be balanced anymore. The circular constant energy surfaces will be deformed to elliptic constant energy surfaces the occupation of which is only set by their (constant) energy. In this way a net current arises from the velocity imbalance.

In the following we will use our moving frame picture based on Hamiltonian~\eqref{tiltEM.eqn} to explain features of the experiment in Ref.~\cite{Ogawa2016} and~\cite{Kuroda2016}. Our theory further predicts that in second order in $\bs\zeta$, interesting polarization-dependence and a novel $\tilde B$-induced Drude peak appears. 

\section{Photo-current in moving frame theory}
Equipped with moving frame picture let us explain essential features of the photo-current experiment in Ref.~\cite{Ogawa2016} and~\cite{Kuroda2016}. The response of the helical electron system in the surface is controlled by an \emph{effective} magnetic field $\tilde B$ that is sum of externally applied $\vec B$ and a term $\vec m$ arising from the Cr magnetic moments. The boost velocity is given by joint effect of these two terms and a possibly non-zero intrinsic $\vec\zeta^{(0)}$. The latter can be either extrinsically induced by external influence~\cite{Motavassal2021}  or can be due to a built-in nematicity imposed on Dirac electrons~\cite{Yekta2023}. 

(1) {\em Transversality of current:} The first and foremost aspect of the observed photo-current indicated in Fig.~1b of Ref.~\cite{Ogawa2016} is that when in-plane magnetic field is applied in $y$ direction, the photo-current is collected in $x$ direction. This immediately follows from Eq.~\eqref{zetaB.eqn}. This equation further predicts that if the in-plane field is applied along the $y$ direction, the corresponding current will be collected in the $-x$ direction. This distinguishes our mechanism from a generic theory of photogalvanic effect for reduced symmetry at the surface~\cite{Belinicher1980,Olbrich2014}. 
Note that Eq.~\eqref{zetaB.eqn} the PH asymmetry $m_*^{-1}$ plays the all-important role of creating \emph{velocity asymmetry}. Hence within our theory, the photo-current can be enhanced in more PH asymmetric (larger $m_*^{-1}$) bands. 

(2) {\em Linearity in low-fields:} This can be immediately seen in our theory as follows: The photo-excitation places an electron on the conduction surface state.  In the absence of boost, each state $+\vec p$ and its time reversed partner $-\vec p$ having opposite group velocities,  cancel each other's effect. Hence the optical excitation will not be able to produce a net current. But in a moving frame there will be a net (group) velocity and hence current imbalance of $2v\vec\zeta$. The approximate linear dependence of the photo-current in Ref.~\cite{Ogawa2016} can be understood by ignoring the intrinsic $\tau_z\vec\zeta^{(0)}$: Within a ballistic picture, one immediately obtains that the induced photo-current is proportional to $\varepsilon_{ab}\tilde B_b$.

(3) \emph{Zero-field photo-current:} As can be seen in Fig. 2b of Ref.~\cite{Ogawa2016}, even at zero applied $B_y$ there is a small but non-zero photo-current. Within our theory, this can be naturally attributed to an intrinsic tilt $\bs\zeta^{(0)}$. A similar intrinsic tilt in the parent non-magnetic compound is clearly visible in the photo-emission data of Ref.~\cite{Kuroda2016}. Within this picture, the contribution of the intrinsic tilt to the photo-current in Ref.~\cite{Ogawa2016} can be on the scale of $\sim 20\%$ of the photo-current induced by $B_y=5$T. The presence of $\tau_z$ in our theory immediately implies that probing the other surface of TI slab will give the opposite zero-field photo-current. The physical cause of non-zero $\vec\zeta^{(0)}$ is a separate question. It could arise from an externally (or perhaps in interacting systems, spontaneously) induced nematicity of Dirac electrons whereby two opposite directions in space are preferred over the other directions~\cite{Motavassal2021}. 

(4) \emph{Dependence on the polarization:} The polarization independence in general grounds is expected when the surface Dirac cones enjoy a rotational invariance~\cite{Hosur2011}. In general the polarization of light responds to anisotroppic environments~\cite{Tudor2016}. Therefore the lack of dependence of the photo-current on polarization in Ref.~\cite{Ogawa2016} can only be understood when the total tilt $\vec\zeta$ is negligible~\cite{Junck2013}.  To see how can polarization dependence arise in moving frame, let us start with the corresponding metric~\eqref{metriccontra.eqn} for which the Kubo formula for undoped Dirac electrons gives
\begin{equation}
    \Pi^{\mu\nu}(q)=\pi(q)\left[q^2 g^{\mu\nu}-q^\mu q^\nu\right], ~\pi(q)=-\frac{1}{16\sqrt{q^2}},
\end{equation}
where $q_\mu=(-\omega,\vec q)$ is the three-momentum and $q^2=q^\mu q_\mu=\bs q^2-(\omega-\bs\zeta.\bs q)^2$~\cite{JalaliMola2019}, reflecting that the underlying spacetime structure is imprinted in the optical response as well. The above response function satisfies the Ward identity $q_\mu \Pi^{\mu\nu}=0$ guaranteeing that the photon remains massless. If for the time-being we ignore the dependence~\eqref{zetaB.eqn} of chemical potential on the $\tilde B$, the above expression gives a polarization dependent current as follows: Assume that the Cartesian coordinate $1$ denotes  the direction of polarization and the direction transverse to polarization in the $(x,y)$ plane is denoted by $2$. Then the conductivity tensor derived from the above $\langle j^\mu j^\nu \rangle$ expression for the optical processes with $q_\mu=(-\omega,\vec 0)$ is~\cite{BruusFlensberg}
\begin{equation}
    \sigma^{ab}=\frac{e^2}{16\hbar}g^{ab}=\frac{e^2}{16\hbar}\left[\delta^{ab}-\zeta^a\zeta^b\right]
\end{equation}
where $\zeta^1$ is the component of $\bs\zeta$ along the polarization. In the absence of tilt, the above expression reduces to the diagonal conductivity of graphene type~\cite{Carbotte2007,Nair2008,Kuzmenko2008}. Denoting the angle between the polarization of the incident light (direction $1$) and the direction of $\vec \zeta$ by $\theta$, the conductivity tensor at zero chemical potential will become
\begin{equation}
    \sigma_{\mu=0}=\frac{e^2}{16\hbar}\begin{pmatrix}
    1-\zeta^2\sin^2\theta & \zeta^2\sin\theta\cos\theta\\
    \zeta^2\sin\theta\cos\theta & 1-\zeta^2\cos^2\theta
    \end{pmatrix}. 
    \label{SpacetimeConductivity.eqn}
\end{equation}
The above expression is frequency independent and suggests that the dependence on polarization appears at the second order of the total tilting parameter $\vec\zeta=\vec\zeta^{(0)}+\vec\zeta^{(1)}$. This gives a term of the form $\zeta^{(0)}\zeta^{(0)}$,  $\zeta^{(0)}\zeta^{(1)}=\zeta^{(0)}\tilde B$ and a second order term of the form $\zeta^{(1)}\zeta^{(1)}\propto \tilde B^2$. Hence the approximate independence on polarization only holds for small $\tilde B$ and small $\zeta^{(0)}$. Carefully \hlt{re-visiting the experiment by} looking for effect that are second order in the above small quantities will reveal polarization dependence of the above form. 

(5) \emph{Second order Drude weight:} At the limit of extremely low-energy photons $\omega\to 0$, due to the moving frame velocity, the Drude weight will be accessible in the photo-current which is given by~\cite{JalaliMola2021}
\begin{equation}
  \frac{e^2\mu \delta(\hbar\omega)}{4\hbar\zeta^2}[1-\sqrt{1-\zeta^2}]\approx\frac{e^2}{16\hbar}\frac{(\mu_B\tilde B)^2}{m_*v^2}\delta(\hbar\omega).
\end{equation}
The above formula offers two advantages: (i) It is a unique prediction of moving frame theory and the observation of the above Drude weight will be additional support of our scenario. (ii) The measurement of the Drude weight at $T>T_c$ in units of $\sigma_0=e^2/(16\hbar)$ as a function of $\tilde B=B_y$ can be used to \hlt{directly determine} the coefficient $1/(m_*v^2)$ of the PH symmetry breaking. 

(6) \emph{Magnetic field induced \hlt{gating}:}
In our theory, the chemical potential itself depends on the tilting parameter~\eqref{zetaB.eqn}. 
The conductivity tensor for non-zero $\mu$ is given by~\cite{Suzumura2014,JalaliMola2021,Mojarro2023} 
\begin{equation}
    \sigma^{11}=\frac{e^2}{16\pi\hbar}\left\{
    \begin{array}{l}
        \theta_X-\cos(2\theta)\sin(2\theta_X)\Theta(1-X^2) \\
        \pi \Theta(-X-1)
    \end{array}    
         \right.
\end{equation}
where $X=(2\mu-\hbar\omega)/\hbar\omega\zeta$ defines $\theta_X=\arccos{X}$ and  $\theta$ as before is the angle between the $\vec\zeta$ and the polarization of the incident light~\cite{Suzumura2014,Verma2017,JalaliMola2021}. The expression for $\sigma^{22}$ can be obtained by replacing $\theta\to\theta+\pi/2$. These corrections arise in the range $|X|<1$ or equivalently 
\begin{equation}
    \frac{2\mu}{1+\zeta}<\hbar\omega<\frac{2\mu}{1-\zeta}.
\end{equation}
The upper and lower limit of the above inequality define twice the aphelion and perihelion of the constant energy ellipse at energy $\mu$. The width of the above interval is proportional to the eccentricity of the constant energy ellipse multiplied by the chemical potential. Since in the present problem the magnetization induced $\mu$~\eqref{zetaB.eqn} itself is proportional to $\tilde B^2$, the energy range in which the above correction appear will have a second order term $\propto \zeta^{(0)}\tilde B^2$ (as well as a third order term $\propto \tilde B^3$) contribution. Therefore the corrections arising from the $\tilde B$-induced chemical potential~\eqref{zetaB.eqn} will induce a deviation from universal conductivity of Dirac electrons over a small energy range around $2\tilde\mu\propto\tilde B^2\hlt{\sim (\zeta^{(1)})^2}$. The photon energies reported in Fig.~2c of Ref.~\cite{Ogawa2016} are above $100$ meV. Observation of the above Fermi level effects requires lower energy photons. \hlt{The above gating effect that relies on $\tilde B$ (an hence on $\zeta$) can give rise to an {\em equilibrium} current away from the transient photo-excited peak. In the following we use this effect can be used to estimate the tilt-parameter in a self-consistent way. }

\section{Discussions}
\hlt{Let us the data in Ref.~\cite{Ogawa2016} to estimate the value of tilt that is induced by external magnetic influence. The current density per electron that arises from moving frame picture is $2e\zeta v$ where the Fermi velocity is $v=4.28\times 10^{5}$ ms$^{-1}$ and $e=1.6\times 10^{-19}$C. The length of the sample used in the experiment along which the current is carried is $L_x=5.6\times 10^{-3}$m. The thickness $L_z=2\times 10^{-9}$m and the lateral dimension drop out from the end estimate of $\zeta$ as follows. To estimate the total number of electrons, since we do not have enough data to estimate the number of photo-excited electrons, we rely on the magnetic-field-induced gating.  
This leads to a background current (away from the {\em transient} photo-excitation peak) that in the present experiment is on the scale of $10^{-7}$A. Our theory at $\tilde B=3T$ gives $\tilde\mu=g\zeta \mu_B\tilde B/2=\zeta g 17.34\times 10^{-5}$ eV. Since the above energy scale is smaller than the thermal energy scale at a typical $20$K of the experiment, assuming a circular Fermi surface, the total number $N$ of the electrons will be given by 
$$
   N=\frac{1}{4\pi}\left[\frac{g\tilde\mu\zeta}{\hbar v}\times 10^{-3}\mbox{m}\right]^2 \approx 3\times 10^{4}(\zeta g)^2,
$$
where we have used the laser spot size $1\times 10^{-3}$m. 
From this a three dimensional current density $j=2\zeta e v N(L_xL_yL_z)^{-1}$ passing through the cross section $L_y L_z$ gives a current
that depends only on $L_x$. Using the g-factor of $g=19.4$~\cite{Wolos2016} for the in-plane fields the current will be given by 
$$
   I=24.61\times 10^{-6}\zeta^3 \mbox{A}\approx 10^{-6}\mbox{A}
$$
that gives $\zeta^{(1)}\sim 0.34$. This is comparable to the tilt values of borophene compounds~\cite{Yekta2023}. A similar tilting of $\sim 0.5$ can be can be obtained from the photoemission data in the spin-orbit coupled Dirac cones of transition metals induced by a magnetic layer. Our estimate of $\zeta$ based on the above gating effects can be larger for larger fields. The intrinsic tilt $\zeta^{(0)}$ being $\sim 20\%$ of the induced tilt is expected to be on the scale of $\zeta^{(0)}\sim 0.08$.  
}

Our theory explains both intrinsic and $\tilde B$-induced zero-bias photo-current on the surface of topological insulators. We further predict nonlinear effects in the form of  polarization-dependence and a non-trivial Drude weight, as well as deviations from universal optical absorption of Dirac electrons in $\omega\to 0$ limit. Both effects can be used to infer information about the breaking of PH symmetry that is represented by $m_*^{-1}$ on which our theory is built. The above additional predictions rely on the  covariance of the optical response a generalized Minkowski spacetime~\cite{JalaliMola2019}. The above covariance \hlt{allows for the propagation of transverse electric (TE)
on the surface of TIs that solely arises from non-zero $\zeta$~\cite{JalaliMola2020}. Since the TE modes can not propagate in other conducting media, the magnetization-enabled tilt can be used as a switch for the TE modes.}

    Our theory also accounts for a similar transient photo-current on the surface of the parent \emph{non-magnetized} Sb$_2$Te$_3$~\cite{Kuroda2016}. This has been attributed to "asymmetry between the transient population of opposite parallel momenta"~\cite{Kuroda2016}. However, as can be seen in Fig. 1(f) of Ref.~\cite{Kuroda2016}, even without the asymmetric population, the photoemission from the photoexcited bands in this study show a clear \emph{intrinsic} velocity asymmetry. This indicates a possible \emph{intrinsic tilt} \hlt{$\zeta^{(0)}$} even without the application of in-plane magnetic field in the surface Dirac cone. In a similar Sb doped system (Bi$_{1-x}$Sb$_x$)$_2$Te$_3$, a systematic dependence of the photo-current to the angle of incidence of the radiation is dubbed photon-drag effect meaning that the momentum $\vec q_{||}$ of the light is transferred to electrons to cause the photo-current~\cite{Ganichev2016}. However, again a substantial intrinsic tilting is clearly visible as velocity imbalance along the $x$ direction~\cite{Ganichev2016}. Within our theory, the contribution of the intrinsic tilt $\vec\zeta^{(0)}$ can be separated from the photon-drag effect by illuminating the other surface of the TI where the intrinsic tilt must reverse its sign. 
Such an intrinsic tilting can also be seen as non-zero photo-current at zero magnetic field in Fig. 2c of Ref.~\cite{Ogawa2016}. In fact the data in this figure suggest that the zero-field contribution to the photo-current is comparable to the photo-current induced by a file of $B_y\sim 1$ T. The subtraction of photo-currents at fields $B_y=5$ T and $B_y=-5$ T in Fig. 3c of Ref.~\cite{Ogawa2016} amounts to subtraction of the background $\zeta^{(0)}$ effect, thereby giving a perfect coincidence between the \emph{antisymmetrized photocurrent} and the magnetization in Fig. 3c. 

\section{Outlook and further possibilities}
\hlt{The fact that moderate magnetic fields of few Tesla are able to induce tilting  values of few tens of percent means that these fields can induce a moving frame velocities that are significant fraction of the Fermi velocity that sets the upper limit of speeds $v$. This resembles a situation where a body moves at e.g. $0.34$ of the speed of light. As such, our synthetic spacetime setup in spin-orbit coupled 2D Dirac cones that arises from our moving frame interpretation allows to design a number of experiments to investigate certain (even strong)} gravitational effects on the table-top:
(i) The extrinsically induced tilt $\bs\zeta$ depends on the joint effect of $B_y$ and the magnetization. The critically enhanced fluctuations of $\vec m$ right above the Curie temperature can be employed to produce larger photo-currents and therefore better detectors of infra-red radiation~\cite{Zhang2020Photodetector,Rogalski2023}. \hlt{Apart from such a device application} this implies that inhomogenously heated sample would imply inhomogenously tilted Dirac cone, and therefore a non-trivial fabric for the emergent spacetime. In particular, a temperature gradient below $T_c$ would imprint a tilt profile with non-zero $\vec\nabla.\vec\zeta$, or a domain wall configuration of $\vec m$ would imply a non-zero $\vec\nabla\times\vec \zeta$. The latter would be a "gravitomagnetic field" \emph{that points along $z$-direction}~\cite{Ryder2009} by simulating a metric~\eqref{metriccovar.eqn} that corresponds to a rotating gravitational source. This can be a fascinating opportunity offered by spin-orbit coupled Dirac cone of the surface of TIs. 
(ii) A magnetic super-structure on the surface of a topological insulator~\cite{vonOppen2014} arranged with wave vector $\vec{q}=(q,0)$ can be used to synthesize a background geometry with gravitational density wave (akin to the familiar spin and charge density wave states~\cite{Gruener2018}). The spin polarization of the stripes can be aligned with an external in-plane $\vec B$ field to generate $\zeta_{x(y)}=\zeta^{(0)}_{x(y)}+\zeta_{x(y)}^{\rm max}\cos(qx)$ profile that can be used to investigate gravitational grating on the tabletop~\cite{Rahvar2018}. In a more generic setting, any form of spatially inhomogeneous magnetic field texture, directly imprints the corresponding texture on the fabric of the spacetime~\eqref{metriccovar.eqn} and would allow to study effects of fluctuations and disorder on the fabric of spacetime that is not easy to measure for the spacetime in the cosmos. (iii) Torsion in condensed matter systems can naturally arise from defects of underlying lattice~\cite{Katanaev2021}. This can be combined with the magnetic manipulation of the Dirac cone in order to synthesize solid-state spacetimes with non-zero torsion~\cite{Liang2020}.  Note that the present spacetime geometry is distinct from quantum geometry~\cite{Ma2021,Ahn2020} that deals with distance in the space of wave-functions rather than the spacetime. 

The ability to influence the properties of synthetic spacetime by magnetic fields on the surface Dirac cones rises hopes to shed light on the role of galactic and extragalactic magnetic fields on the fabric of spacetime~\cite{Tsagas2001}. 

{\bf Acknowledgements:} This work was supported by the Alexander von Humboldt foundation. I am grateful to Eva Pavarini and David Divincenzo for hospitality. I wish to thank Ying-Jiun Chen and Kristof Moors for their comments on the manuscript and Eduar Zsurka for useful discussions.

\bibliography{Refs}



\end{document}